\documentclass[12pt,a4paper]{article}
\usepackage{amsmath}
\usepackage{graphicx}
\usepackage{amsfonts}
\usepackage{amssymb}
\usepackage{epsfig}

\newcommand{\nc}{\newcommand}
\nc{\renc}{\renewcommand}

\nc{\half}{{\textstyle{1\over2}}}
\nc{\etal}{\mbox{\it et al. }} \nc{\ie}{{\it i.e.}} \nc{\eg}{{\it
e.g.}}

\renc{\thefootnote}{\arabic{footnote}} \nc{\capt}[1]{{\bf
Figure.} {\small\sl #1}}

\nc{\eqs}[2]{\mbox{Eqs.~(\ref{#1},\,\ref{#2})}}
\nc{\eq}[1]{\mbox{Eq.~(\ref{#1})}}

\nc{\figs}[2]{\mbox{Figs.~(\ref{#1},\,\ref{#2})}}
\nc{\fig}[1]{\mbox{Fig~.(\ref{#1})}}

\nc{\mtag}[1]{\label{#1} \mbox{\marginpar{{\footnotesize #1}}}}
\newlength{\overeqskip}
\newlength{\undereqskip}
\nc{\be}[1]{\begin{equation} \mbox{$\label{#1}$}}
\nc{\bea}[1]{\begin{eqnarray} \mbox{$\label{#1}$}}
\nc{\Section}[2]{\section{#2}\label{#1}}
\nc{\Bibitem}[1]{\bibitem{#1}} \nc{\Label}[1]{\label{#1}}

\nc{\eea}{\vspace{\undereqskip}\end{eqnarray}}
\nc{\ee}{\vspace{\undereqskip}\end{equation}}
\nc{\bdm}{\begin{displaymath}} \nc{\edm}{\end{displaymath}}
\nc{\dpsty}{\displaystyle} \nc{\bc}{\begin{center}}
\nc{\ec}{\end{center}} \nc{\ba}{\begin{array}}
\nc{\ea}{\end{array}} \nc{\bab}{\begin{abstract}}
\nc{\eab}{\end{abstract}} \nc{\btab}{\begin{tabular}}
\nc{\etab}{\end{tabular}} \nc{\bit}{\begin{itemize}}
\nc{\eit}{\end{itemize}} \nc{\ben}{\begin{enumerate}}
\nc{\een}{\end{enumerate}} \nc{\bfig}{\begin{figure}}
\nc{\efig}{\end{figure}}
\nc{\arreq}{&\!=\!&} \nc{\arrmi}{&\!-\!&} \nc{\arrpl}{&\!+\!&}
\nc{\arrap}{&\!\!\!\approx\!\!\!&} \nc{\non}{\nonumber\\*}

\def\lsim{\; \raise0.3ex\hbox{$<$\kern-0.75em
      \raise-1.1ex\hbox{$\sim$}}\; }
\def\gsim{\; \raise0.3ex\hbox{$>$\kern-0.75em
      \raise-1.1ex\hbox{$\sim$}}\; }
\nc{\DOT}{\hspace{-0.08in}{\bf .}\hspace{0.1in}}
\nc{\Laada}{\hbox {$\sqcap$ \kern -1em $\sqcup$}}
\nc\loota{{\scriptstyle\sqcap\kern-0.55em\hbox{$\scriptstyle\sqcup$}}}
\nc\Loota{{\sqcap\kern-0.65em\hbox{$\sqcup$}}} \nc\laada{\Loota}
\nc{\qed}{\hskip 3em \hbox{\BOX} \vskip 2ex}

\nc{\real}{{\rm I \! R}} \nc{\Z}{{\sf Z \!\!\! Z}}
\nc{\complex}{{\rm C\!\!\! {\sf I}\,\,}}
\def\bigid{\leavevmode\hbox{\small1\kern-3.8pt\normalsize1}}
\def\id{\leavevmode\hbox{\small1\kern-3.3pt\normalsize1}}
\nc{\slask}{\!\!\!/} \nc{\bis}{{\prime\prime}} \nc{\pa}{\partial}
\nc{\na}{\nabla} \nc{\ra}{\rangle} \nc{\la}{\langle}
\nc{\goto}{\rightarrow} \nc{\swap}{\leftrightarrow}

\nc{\EE}[1]{ \mbox{$\cdot10^{#1}$} } \nc{\abs}[1]{\left|#1\right|}
\nc{\at}[2]{\left.#1\right|_{#2}} \nc{\norm}[1]{\|#1\|}
\nc{\abscut}[2]{\Abs{#1}_{\scriptscriptstyle#2}}
\nc{\vek}[1]{{\rm\bf #1}} \nc{\integral}[2]{\int\limits_{#1}^{#2}}
\nc{\inv}[1]{\frac{1}{#1}} \nc{\dd}[2]{{{\partial
#1}\over{\partial #2}}} \nc{\ddd}[2]{{{{\partial}^2
#1}\over{\partial {#2}^2}}} \nc{\dddd}[3]{{{{\partial}^2 #1}\over
        {\partial #2 \partial #3}}}
\nc{\dder}[2]{{{d #1}\over{d #2}}} \nc{\ddder}[2]{{{d^2
#1}\over{d {#2}^2}}} \nc{\dddder}[3]{{d^2 #1}\over
        {d #2 d #3}}
\nc{\dx}[1]{d\,^{#1}x} \nc{\dy}[1]{d\,^{#1}y}
\nc{\dz}[1]{d\,^{#1}z} \nc{\dl}[1]{\frac{d\,^{#1}l}{(2\pi)^{#1}}}
\nc{\dk}[1]{\frac{d\,^{#1}k}{(2\pi)^{#1}}}
\nc{\dq}[1]{\frac{d\,^{#1}q}{(2\pi)^{#1}}}

\nc{\cc}{\mbox{$c.c.$ }} \nc{\hc}{\mbox{$h.c.$ }} \nc{\cf}{cf.\ }
\nc{\erfc}{{\rm erfc}} \nc{\Tr}{{\rm Tr\,}} \nc{\tr}{{\rm tr\,}}
\nc{\pol}{{\rm pol}} \nc{\sign}{{\rm sign}} \nc{\bfT}{{\bf T }}

\nc{\cA}{{\cal A}} \nc{\cB}{{\cal B}} \nc{\cD}{{\cal D}}
\nc{\cE}{{\cal E}} \nc{\cG}{{\cal G}} \nc{\cH}{{\cal H}}
\nc{\cL}{{\cal L}} \nc{\cO}{{\cal O}} \nc{\cT}{{\cal T}}
\nc{\cN}{{\cal N}}
\nc{\rvac}[1]{|{\cal O}#1\rangle} \nc{\lvac}[1]{\langle{\cal
O}#1|} \nc{\rvacb}[1]{|{\cal O}_\beta #1\rangle}
\nc{\lvacb}[1]{\langle{\cal O}_\beta #1 |} \nc{\bb}{\bar{\beta}}
\nc{\bt}{\tilde{\beta}} \nc{\ctH}{\tilde{\cal H}}
\nc{\chH}{\hat{\cal H}}
\nc{\1}{\aa} \nc{\2}{\"{a}} \nc{\3}{\"{o}} \nc{\4}{\AA}
\nc{\5}{\"{A}} \nc{\6}{\"{O}}
\nc{\al}{\alpha} \nc{\g}{\gamma} \nc{\Del}{\Delta}
\nc{\e}{\epsilon} \nc{\eps}{\epsilon} \nc{\lam}{\lambda}
\nc{\om}{\omega} \nc{\Om}{\Omega} \nc{\ve}{\varepsilon}
\nc{\mn}{{\mu\nu}}
\nc{\vp}{\varphi}

\nc{\aap}[3]{{\it  Astron.\ Astrophys.\ }{{\bf #1} {(#2)} {#3}}}
\nc{\advp}[3]{{\it  Adv.\ in\ Phys.\ }{{\bf #1} {(#2)} {#3}}}
\nc{\annp}[3]{{\it  Ann.\ Phys.\ (N.Y.)\ }{{\bf #1} {(#2)} {#3}}}
\nc{\apl}[3]{{\it  Appl. Phys. Lett. }{{\bf #1} {(#2)} {#3}}}
\nc{\apj}[3]{{\it  Ap.\ J.\ }{{\bf #1} {(#2)} {#3}}}
\nc{\apjl}[3]{{\it  Ap.\ J.\ Lett.\ }{{\bf #1} {(#2)} {#3}}}
\nc{\app}[3]{{\it Astropart.\ Phys.\ }{{\bf #1} {(#2)} {#3}}}
\nc{\cmp}[3]{{\it  Comm.\ Math.\ Phys.\ }{{ \bf #1} {(#2)} {#3}}}
\nc{\cqg}[3]{{\it  Class.\ Quant.\ Grav.\ }{{\bf #1} {(#2)} {#3}}}
\nc{\epj}[3]{{\it  Eur.\ Phys.\ J.\ }{{\bf #1} {(#2)} {#3}}}
\nc{\epl}[3]{{\it  Europhys.\ Lett.\ }{{\bf #1} {(#2)} {#3}}}
\nc{\ijmp}[3]{{\it Int.\ J.\ Mod.\ Phys.\ }{{\bf #1} {(#2)} {#3}}}
\nc{\ijtp}[3]{{\it Int.\ J.\ Theor.\ Phys.\ }{{\bf #1} {(#2)}
{#3}}} \nc{\jmp}[3]{{\it  J.\ Math.\ Phys.\ }{{ \bf #1} {(#2)}
{#3}}} \nc{\jpa}[3]{{\it  J.\ Phys.\ A\ }{{\bf #1} {(#2)} {#3}}}
\nc{\jpc}[3]{{\it  J.\ Phys.\ C\ }{{\bf #1} {(#2)} {#3}}}
\nc{\jap}[3]{{\it J.\ Appl.\ Phys.\ }{{\bf #1} {(#2)} {#3}}}
\nc{\jpsj}[3]{{\it J.\ Phys.\ Soc.\ Japan\ }{{\bf #1} {(#2)}
{#3}}} \nc{\lmp}[3]{{\it Lett.\ Math.\ Phys.\ }{{\bf #1} {(#2)}
{#3}}} \nc{\mpl}[3]{{\it  Mod.\ Phys.\ Lett.\ }{{\bf #1} {(#2)}
{#3}}} \nc{\ncim}[3]{{\it  Nuov.\ Cim.\ }{{\bf #1} {(#2)} {#3}}}
\nc{\np}[3]{{\it  Nucl.\ Phys.\ }{{\bf #1} {(#2)} {#3}}}
\nc{\pr}[3]{{\it Phys.\ Rev.\ }{{\bf #1} {(#2)} {#3}}}
\nc{\pra}[3]{{\it  Phys.\ Rev.\ A\ }{{\bf #1} {(#2)} {#3}}}
\nc{\prb}[3]{{\it  Phys.\ Rev.\ B\ }{{{\bf #1} {(#2)} {#3}}}}
\nc{\prc}[3]{{\it  Phys.\ Rev.\ C\ }{{\bf #1} {(#2)} {#3}}}
\nc{\prd}[3]{{\it  Phys.\ Rev.\ D\ }{{\bf #1} {(#2)} {#3}}}
\nc{\prl}[3]{{\it Phys.\ Rev.\ Lett.\ }{{\bf #1} {(#2)} {#3}}}
\nc{\pl}[3]{{\it  Phys.\ Lett.\ }{{\bf #1} {(#2)} {#3}}}
\nc{\prep}[3]{{\it Phys.\ Rep.\ }{{\bf #1} {(#2)} {#3}}}
\nc{\prsl}[3]{{\it Proc.\ R.\ Soc.\ London\ }{{\bf #1} {(#2)}
{#3}}} \nc{\ptp}[3]{{\it  Prog.\ Theor.\ Phys.\ }{{\bf #1} {(#2)}
{#3}}} \nc{\ptps}[3]{{\it  Prog\ Theor.\ Phys.\ suppl.\ }{{\bf
#1} {(#2)} {#3}}} \nc{\physa}[3]{{\it  Physica\ A\ }{{\bf #1}
{(#2)} {#3}}} \nc{\physb}[3]{{\it  Physica\ B\ }{{\bf #1} {(#2)}
{#3}}} \nc{\phys}[3]{{\it Physica\ }{{\bf #1} {(#2)} {#3}}}
\nc{\rmp}[3]{{\it  Rev.\ Mod.\ Phys.\ }{{\bf #1} {(#2)} {#3}}}
\nc{\rpp}[3]{{\it Rep.\ Prog.\ Phys.\ }{{\bf #1} {(#2)} {#3}}}
\nc{\sjnp}[3]{{\it Sov.\ J.\ Nucl.\ Phys.\ }{{\bf #1} {(#2)}
{#3}}} \nc{\spjetp}[3]{{\it Sov.\ Phys.\ JETP\ }{{\bf #1} {(#2)}
{#3}}} \nc{\yf}[3]{{\it Yad.\ Fiz.\ }{{\bf #1} {(#2)} {#3}}}
\nc{\zetp}[3]{{\it Zh.\ Eksp.\ Teor.\ Fiz.\  }{{\bf #1}  {(#2)}
{#3}}} \nc{\zp}[3]{{\it Z.\ Phys.\ }{{\bf #1} {(#2)} {#3}}}
\nc{\ibid}[3]{{\sl ibid.\ }{{\bf #1} {(#2)} {#3}}}
\nc{\rf}[1]{(\ref{#1})} \nc{\nn}{\nonumber \\*} \nc{\bfB}{\bf{B}}
\nc{\bfv}{\bf{v}} \nc{\bfx}{\bf{x}} \nc{\bfy}{\bf{y}}
\nc{\vx}{\vec{x}} \nc{\vy}{\vec{y}} \nc{\oB}{\overline{B}}
\nc{\oI}{\overline{I}} \nc{\oR}{\overline{R}}
\nc{\rar}{\rightarrow} \nc{\ti}{\times} \nc{\slsh}{\hskip-5pt/}
\nc{\sm}{Standard~Model~} \nc{\MP}{M_{\rm Pl}} \nc{\tp}{t_{\rm
Pl}} \nc{\ave}{\bar{E}}

\nc{\eff}{{\rm eff}} \nc{\kk}{\vek{k}} \nc{\pp}{{\rm p}}
\nc{\ga}{g_{a\gamma}} \nc{\vv}{\\} \nc{\eee}{{\bf E}}
\nc{\bbb}{{\bf B}} \nc{\qcd}{T_{\rm QCD}} \nc{\G}{\rm \ G}
\begin{document}
\begin{titlepage}
\begin{flushright}
CFNUL/01-1 \\
hep-ph/0107235
\end{flushright}
\vspace*{0.2cm}
\begin{center}
{\Large \bf Effects of periodic matter in kaon regeneration} \\
\vspace{0.8cm}
{\large Evgeny Akhmedov$^{a,1}$, Augusto Barroso$^{b,2}$ and
Petteri Ker\"anen$^{b,3}$ } \\
\vspace{0.3cm}
{\em $^{a}$ Institut f{\"u}r Theoretische Physik T30, Physik
Department \\ Technische Universit{\"a}t
M{\"u}nchen, James-Franck-Stra{\ss}e, \\
D-85748 Garching b. M{\"u}nchen, Germany} \\

\vspace{0.2cm}
{\em $^{b}$ Centro de F{\'{\i}}sica Te\'orica e
Computacional,
Faculdade de Ci\^encias\\ Universidade de Lisboa,
Av. Prof. Gama Pinto 2\\ P-1649-003 Lisboa, Portugal  }
\end{center}
\begin{abstract}
\noindent We study the effects of periodic matter in kaon regeneration,
motivated by the possibility of parametric resonance in neutrino
oscillations. The large imaginary parts of the forward kaon-nucleon
scattering amplitudes and the decay width difference $\Delta\Gamma$
prevent a sizable enhancement of the $K_L\rightarrow K_S$ transition
probability. However, some interesting effects can be produced
using regenerators made of alternating layers of two different
materials. Despite the fact that the regenerator has a fixed
length one can obtain different values for the probability distribution
of the $K_L$ decay into a final state. Using a two-arm regenerator set up
it is possible to measure the imaginary parts of the $K^0(\bar{K}^0)$-nucleon 
scattering amplitudes in the correlated decays of the $\phi$-resonance.
Combining the data of the single-arm regenerator experiments with direct
and reverse orders of the matter layers in the regenerator one can
independently measure the CP violating parameter $\delta$.
\end{abstract}

\vskip10pt
\noindent
{\it PACS:} 11.30.Er; 13.75.Jz

\noindent {\it Keywords:} kaon oscillations; regeneration
\vfil
\noindent
\footnoterule  {\small $^1$On leave from
National Research Centre Kurchatov Institute, Moscow 123182, Russia.
\vskip-1pt\noindent}{\small
akhmedov@physik.tu-muenchen.de\vskip-1pt\noindent} {\small
$^{2}$barroso@.cii.fc.ul.pt\vskip-1pt\noindent} {\small
$^{3}$keranen@cii.fc.ul.pt\vskip-1pt\noindent}

\thispagestyle{empty}
\end{titlepage}
\setcounter{page}{1}

Recently, there has been a renewed interest \cite{renewed} in the possibility
of parametric resonance in neutrino oscillations in matter suggested
in \cite{ETC,Akh1}. For a neutrino beam propagating in a medium with
periodic density, one can obtain a large probability for the transition
from one flavour state to another, even if the neutrino mixing angles
both in vacuum and in matter are small.

In nature, there are other systems similar to oscillating neutrinos,
in particular the neutral mesons $K^0 - \bar{K}^0$. Hence, it is
interesting to investigate if one can obtain the parametric resonance in
this case. The analogue of the neutrino weak flavour basis are $K^0$ and
$\bar{K^0}$ and the mass eigenstates are $K_L$ and $K_S$. Since the former
states are maximally mixed, it is obvious that one cannot enhance the $K^0
- \bar{K}^0$ transition probability. However, in this case this
is not the relevant question.

Let us assume that we have a neutral kaon beam propagating in
vacuum. After a time $t$ larger than the $K_S$ lifetime
($\tau_S=0.894\times 10^{-10}$~s) the beam is essentially a $K_L$
beam. If this beam traverses a thin slab of material, a small
$K_S$ component will emerge, because $K^0$ and $\bar{K}^0$ have
different scattering amplitudes. This is the well-known
regeneration phenomenon (see, e.g., \cite{BLS}). Assuming that
the beam enters the regenerator at $t=0$ and denoting by
$|K_R(t)\rangle$ the state of the beam when it emerges, we have
\begin{eqnarray}\label{Kdecay}
|K_R(t)\rangle = \langle \tilde{K}_S|T|K_L\rangle |K_S\rangle +
\langle \tilde{K}_L|T|K_L\rangle |K_L\rangle ,
\end{eqnarray}
where $\langle \tilde{K}_{S,L}|T|K_L\rangle$ are the transition
amplitudes in the regenerator, and $\langle \tilde{K}_{S,L}|$ are
the reciprocal states (see Eqs. (\ref{ketKL}) and (\ref{ketKS}) below).
If the regenerator
is a medium with a density that is a periodic function of the coordinate
along the beam direction, we would like to see if it is possible to
enhance the $K_L \rightarrow K_S$ transition amplitude. Our aim in this
letter is to address this question.

Assuming CPT conservation, but not CP conservation, the rest-frame
evolution equation for the $K^0 - \bar{K}^0$ system propagating in a
medium is
\begin{eqnarray}\label{evol}
i \frac{{\rm d}}{{\rm d} t} \left(\begin{array}{c}
 K^0 \\ \bar{K}^0
\end{array} \right)\ =
\left(\begin{array}{cc}
\mu+V   & \frac{p}{q}\frac{\Delta\mu}{2} \\
\frac{q}{p}\frac{\Delta\mu}{2}   & \mu+\bar{V}    \\
\end{array}\right)\
\left(\begin{array}{c}
 K^0 \\  \bar{K}^0
\end{array} \right)\ ,
\end{eqnarray}
where $t$ is the proper time (we follow closely the notation of
ref.~\cite{BLS}). Hence, in vacuum ($V=\bar{V}=0$)
the eigenstates of the Hamiltonian $H_0$ are
\begin{eqnarray}\label{braKL}
|K_L\rangle & = & p |K^0\rangle + q |\bar{K}^0\rangle
\end{eqnarray}
and
\begin{eqnarray}
|K_S\rangle & = & p |K^0\rangle - q |\bar{K}^0\rangle \
\label{braKS},
\end{eqnarray}
with the corresponding eigenvalues
$\mu_{L,S}=m_{L,S}-\frac{i}{2}\Gamma_{L,S}$, $\mu
=(\mu_L+\mu_S)/2$ and $\Delta\mu=\mu_L-\mu_S$.
Since the phase of $p/q$ is of no physical significance, we shall assume
this ratio to be real\footnote{Since the relative phases between $p/q$ and
certain ratios of amplitudes do have physical content, the
phase convention we use here implies a phase convention for the
decay amplitudes as well.}. We write
\begin{equation}
p = \sqrt{\frac{1+\delta}{2}}\,,\qquad
q = \sqrt{\frac{1-\delta}{2}}\,,
\label{qwithdelta}
\end{equation}
where $\delta\simeq 3\times 10^{-3}$ \cite{RPP} is a measure of CP
violation. Since CP is not conserved, the diagonalization of $H_0$  cannot
be accomplished with a unitary transformation. This, in turn, implies the
use of the reciprocal basis~\cite{reciprocalSilva}
\begin{eqnarray}\label{ketKL}
\langle \tilde{K}_L| &= &\frac{1}{2}\, \left(\frac{1}{p} \langle
K^0| +
\frac{1}{q} \langle \bar{K}^0|\right) \ ,\\
\langle \tilde{K}_S| &= &\frac{1}{2}\, \left(\frac{1}{p} \langle
K^0| - \frac{1}{q} \langle \bar{K}^0|\right) \label{ketKS}.
\end{eqnarray}
In a medium with $N_a$ nuclei per unit volume, $V\ (\bar{V})$ is
given in terms of the forward scattering amplitude $f(0)\
(\bar{f}(0))$ for a $K^0 \ (\bar{K}^0)$
beam \cite{BLS}, i.e.
\begin{eqnarray}
V=-\frac{2\pi N_a}{m}f(0) \ ,
\end{eqnarray}
with the average kaon mass $m=2.52$~fm$^{-1}$. The simplest way
to introduce a periodic medium is to consider two different
elements with number densities $N_a$ and $N_b$ positioned one
after the other and to build a regenerator with $\kappa$ layers of
this $ab$ junction. The beam evolution through this multilayer
regenerator can be described in terms of the evolution operator
\begin{eqnarray}
U_{\kappa}=\underbrace{U_b U_a \times \ ... \ \times U_b
U_a}_{\kappa}\, ,
\end{eqnarray}
with
\begin{eqnarray}
U_a&=&\exp\left( -iH_a t_a \right) \label{Ua}
\end{eqnarray}
and
\begin{eqnarray}
U_b&=&\exp\left( -iH_b t_b \right) \label{Ub}\ ,
\end{eqnarray}
where $H_i$ ($i=a,b$) are the Hamiltoneans for layers $a$ or $b$,
given in Eq. (\ref{evol}). Since this is a $2\times 2$ matrix it
is convenient to represent it using the Pauli $\sigma$ matrices.
One can then write
\begin{equation}\label{Ha}
H_a=F_a+\mbox{\boldmath $\sigma$}{\bf E}_a\,,
\end{equation}
where
\begin{eqnarray}
F_a=\mu+\frac{V_a+\bar{V}_a}{2}\, ,
\end{eqnarray}
and ${\bf E}_a$ is a three dimensional vector with components
\begin{eqnarray}\label{Ea1}
E_a^{(1)}&=&\frac{\Delta\mu}{4}\left(\frac{p}{q}+\frac{q}{p}\right)\,
,\\
E_a^{(2)}&=&i\frac{\Delta\mu}{4}\left(\frac{p}{q}-\frac{q}{p}\right)\,
,\label{Ea2}\\
E_a^{(3)}&=&\frac{V_a-\bar{V}_a}{2} \equiv\frac{\Delta V_a}{2} \,
\label{Ea3},
\end{eqnarray}
which are complex numbers. Introducing the complex unit vector
\begin{eqnarray}
{\bf n}_a=\frac{{\bf E}_a}{E_a}\,,\qquad E_a \equiv \sqrt{{\bf
E}_a\cdot{\bf E}_a}\,,
\end{eqnarray}
and
\begin{eqnarray}
\varphi_a=E_a t_a \,,
\end{eqnarray}
one immediately obtains
\begin{eqnarray}\label{Uaopen}
U_a=\exp\left( -iF_at_a \right)\left( \cos\varphi_a
-i\mbox{\boldmath $\sigma$} \cdot {\bf n}_a \sin\varphi_a\right) \, ,
\end{eqnarray}

With the obvious replacements $a\rightarrow b$ one obtains from
Eq.~(\ref{Uaopen}) $U_b$. Then the product $U_bU_a$ is
\begin{eqnarray}\label{UbUa}
U_bU_a=\exp \left[-i\left( F_a t_a + F_b t_b \right)\right]
\left[Y-i\mbox{\boldmath $\sigma$} \cdot{\bf X}\right]\, ,
\end{eqnarray}
with
\begin{eqnarray}\label{Y}
Y & = & \cos\varphi_a \cos\varphi_b - \sin\varphi_a \sin\varphi_b
({\bf n}_a \cdot {\bf n}_b) \ ,\\
{\bf X} & = & \sin\varphi_a \cos\varphi_b {\bf n}_a +
\sin\varphi_b \cos\varphi_a {\bf n}_b - \sin\varphi_a
\sin\varphi_b ({\bf n}_a \times {\bf n}_b ) \ \label{vecX}.
\end{eqnarray}
The vectors that we have introduced have complex components.
However, the dot products, such as ${\bf n}_a\cdot{\bf n}_b$, must
be simply understood as
\begin{eqnarray}
{\bf n}_a\cdot{\bf n}_b=\sum_i^3 n_a^{(i)} n_b^{(i)}\, .
\end{eqnarray}
Notice that the third component of ${\bf n}_a \times {\bf n}_b$
is identically zero. Then Eq.~(\ref{vecX}) shows that $X^{(3)}$
is symmetric with respect to the interchange of $a$ and $b$. On the other
hand, Eq.~(\ref{Ea2}) shows that $n_a^{(2)}$ and $n_b^{(2)}$
vanish in the limit of CP conservation. Then, in this limit,
$X^{(2)}$ is antisymmetric in $a$ and $b$. Furthermore, in the
same approximation the first component of ${\bf n}_a \times
{\bf n}_b$ is also zero. Hence $X^{(1)}$ is symmetric with respect to
the interchange of $a$ and $b$.

A straightforward calculation shows that
$Y^2+{\bf X}\cdot{\bf X}=1$. Then, defining another complex angle
$\Phi$ such that
\begin{eqnarray}
\cos\Phi&=&Y\ ,\\
\sin\Phi&=&\sqrt{{\bf X}^2}\ ,
\end{eqnarray}
it is possible to rewrite Eq. (\ref{UbUa}) in the form
\begin{eqnarray}
U_b U_a=\exp\left[-i\left(F_at_a+F_bt_b \right)\right]\
\exp\left(-i\mbox{\boldmath $\sigma$}\!\cdot\!\hat{\bf X}\,\Phi\right)\,
\end{eqnarray}
with
\begin{eqnarray}
\hat{\bf X}=\frac{{\bf X} }{ \sqrt{{\bf X}^2} }\ .
\end{eqnarray}
This evolution operator is written in the $K^0-\bar{K}^0$-basis.
Denoting $U_{ab}\equiv U_bU_a$, the symmetry properties of $X^{(i)}$
deduced above enables us to obtain
\begin{eqnarray}
\langle K^0 |U_{ab}|K^0 \rangle&=&\langle K^0|U_{ba}|K^0\rangle\, ,\\
\langle \bar{K}^0 |U_{ab}|\bar{K}^0 \rangle &=& \langle
\bar{K}^0 |U_{ba}|\bar{K}^0 \rangle \, ,
\end{eqnarray}
but
\begin{eqnarray}
\langle K^0|U_{ab}|\bar{K}^0 \rangle - \langle
\bar{K}^0 |U_{ba}|K^0\rangle&\propto& \delta \, ,\\
\langle \bar{K}^0 |U_{ab}|K^0\rangle -\langle K^0|U_{ba}|\bar{K}^0
\rangle &\propto & \delta \, ,
\end{eqnarray}
i.e. the difference vanishes if CP is conserved.

Finally, the evolution matrix for the propagation through $\kappa$
$ab$-layers is simply
\begin{eqnarray}
U_\kappa=\exp\left[-i\kappa(F_at_a+F_bt_b)\right]
\left(\cos(\kappa\Phi)-i\mbox{\boldmath $\sigma$}
\cdot\hat{\bf X}\sin(\kappa\Phi)
\right)\ .
\end{eqnarray}
Inserting $U_\kappa$ between the appropriate bra- and ket-vectors
given by Eqs.~(\ref{braKL})-(\ref{braKS}) and
(\ref{ketKL})-(\ref{ketKS}) respectively, one obtains the
$K_L\rightarrow K_L$ and the $K_L\rightarrow K_S$ transition
amplitudes
\begin{eqnarray}\label{AKLKL}
\langle \tilde{K}_L|U_\kappa|K_L\rangle &=& \exp\left[ -i\kappa
\left(F_at_a+F_bt_b \right)\right]\times \nonumber\\
&\times &\left\{\cos(\kappa\Phi)-\frac{i}{2}\sin(\kappa\Phi)
\left( \left[ \frac{p}{q}+\frac{q}{p}\right]\hat{X}_1
+i\left[\frac{p}{q}-\frac{q}{p}\right]\hat{X}_2  \right)\right\}
\end{eqnarray}
and
\begin{eqnarray}\label{AKLKS}
\langle \tilde{K}_S|U_\kappa|K_L\rangle &=& \exp\left[ -i\kappa \left(
F_at_a+F_bt_b \right)\right]\times \nonumber\\
&\times &\frac{i}{2}\sin(\kappa\Phi)\left\{ -2\hat{X}_3 +\left[
\frac{p}{q}-\frac{q}{p}\right]\hat{X}_1
+i\left[\frac{p}{q}+\frac{q}{p}\right] \hat{X}_2 \right\} \, .
\end{eqnarray}

Let us start our discussion with a careful examination of
Eqs.~(\ref{Ea1})-(\ref{Ea3}). The vectors
${\bf E}_a$ and ${\bf E}_b$ have the first components
proportional to $\Delta\mu/2$ and the third components
proportional to $\Delta V_i/2$. These quantities $\Delta\mu$ and
$\Delta V$ play a crucial role in the effect that we are
searching for. On the contrary, the mean values $\mu$ and
$(V_i+\bar{V}_i)/2$ are far less important. Their real parts
disappear when we take the modulus square of the amplitude to
obtain the transition probabilities and their imaginary parts
give the overall damping factors.

As a first approximation, we neglect CP violation. Let us further assume
that $\Delta\mu$ and $\Delta V_i$ are real. A real $\Delta\mu$ means
that $\Delta\Gamma=0$. Although this is not true for the $K$-meson
system, there is no fundamental reason why it could not be so.
Indeed, such a situation occurs closely for the $B^0-\bar{B}^0$
mesons. A real $\Delta V_i$ implies equal imaginary parts for the
$K^0$ and $\bar{K}^0$ forward scattering amplitudes. As it is
well known, this is not the case. This is in contrast with the case
of neutrinos, where the absorption is weak, and to the leading order in
weak interaction the scattering amplitudes are real.

\begin{figure}
 \begin{center}
 \epsfig{file=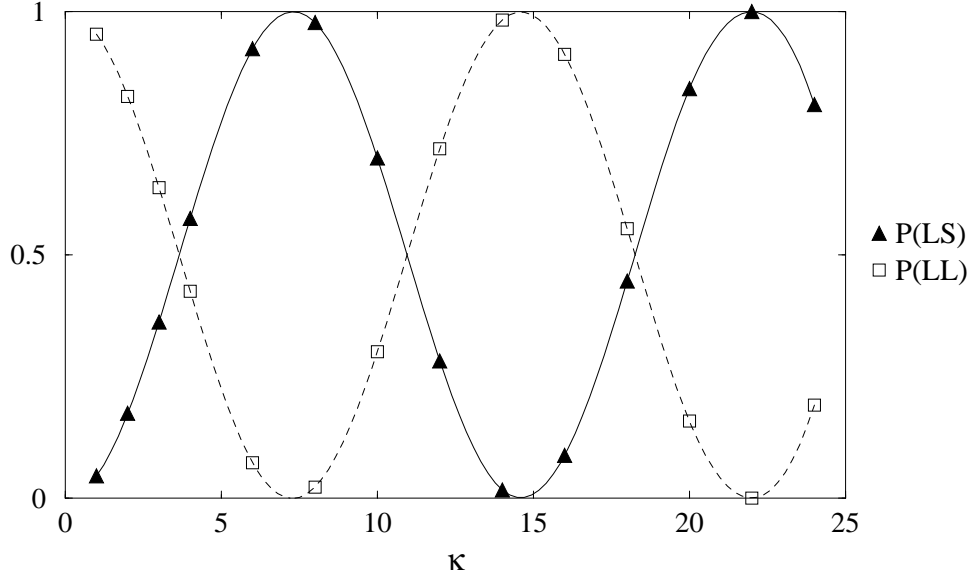,width=8cm,angle=270}
    \caption{$K_L\rightarrow K_S$ and $K_L\rightarrow K_L$ transition
            probabilites $P(LS)$ and $P(LL)$, respectively, with
            ${\rm Im}(V)={\rm Im}(\bar V)=\Delta\Gamma=\Gamma=0$.
            The regenerator is $^{27}{\rm Al}\,-^{184}$W and
            $t_a=59.21\times10^{-11}$~s and $t_b=57.29\times10^{-11}$~s.}
    \label{fig:fig1}
  \end{center}
\end{figure}

Within this unrealistic approximation it is possible to achieve a
parametric resonance in $K_L\leftrightarrow K_S$ transitions in matter.
The parametric resonance condition is $X_3=0$ \cite{rescond}; we shall
consider a particular realization of this condition in which
the times $t_a$ and $t_b$ are chosen such that
$\cos\varphi_a=\cos\varphi_b=0$. Then it follows from
Eq.~(\ref{vecX}) that
\begin{eqnarray}
{\bf X}=\pm({\bf n}_a\times {\bf n}_b)\, .
\end{eqnarray}
As described above, the third component of the cross product
${\bf n}_a \times {\bf n}_b$ is identically zero and if one
neglects CP violation the first component is also zero. In this
approximation Eqs.~(\ref{AKLKL}) and (\ref{AKLKS}) become
\begin{eqnarray}\label{AKLKLresonance}
\langle \tilde{K}_L|U_\kappa|K_L\rangle &=& \exp\left[ -i\kappa \left(
F_at_a+F_bt_b \right)\right]\cos(\kappa\Phi)
\end{eqnarray}
and
\begin{eqnarray}\label{AKLKSresonance}
\langle \tilde{K}_S|U_\kappa|K_L\rangle &=& -\exp\left[ -i\kappa \left(
F_at_a+F_bt_b \right)\right]\sin(\kappa\Phi)\hat{X}_2  \, .
\end{eqnarray}
For an appropriate number of layers, $\kappa$, one can suppress the
$K_L\rightarrow K_L$ probability and, at the same time, enhance
the $K_L\rightarrow K_S$ transition probability. To illustrate
this effect, we plot the $K_L\rightarrow K_S$ transition
probability as a function of $\kappa$ in Fig.~\ref{fig:fig1}. The
calculation was done for a regenerator made of $^{27}$Al and
$^{184}$W and for an initial $K_L$ beam obtained from the decay
of the $\phi$ resonance at rest. The values of the $K^0$ and
$\bar{K}^0$ scattering amplitudes on protons and neutrons
were taken from ref.~\cite{scattamplitudes}. As we have explained,
$\Delta\Gamma$ and the imaginary parts of $\Delta V_i$ were set
equal to zero.

The times $t_a$ and $t_b$ were chosen in such a way that a complete
$K_L\rightarrow K_S$ conversion could be obtained. If we move
away from this resonance condition we still obtain an oscillatory
$K_L\rightarrow K_S$ transition probability $P(K_L\rightarrow
K_S)$ but with a smaller maximal conversion. For instance,
decreasing both $t_a$ and $t_b$ by 17~\% reduces the maximum value
of $P(K_L\rightarrow K_S)$ from 1 to 0.145.

The resonance values of $t_a$ and $t_b$ ($59.210\times 10^{-11}$~s
and $57.289\times 10^{-11}$~s) are a factor of seven or six
larger than $\tau_S$. This by itself is sufficient to explain
that the effect disappears as soon as we introduce the right
values of $\Gamma$ and $\Delta\Gamma$, even with Im$(V)={\rm
Im}(\bar{V})=0$. We have checked that, in this case,
$P(K_L\rightarrow K_S)\simeq 10^{-4}$ for $\kappa=1$ and decreases
slowly with $\kappa$. In addition, if we introduce the correct
values for the imaginary parts of the scattering amplitudes,
$P(K_L\rightarrow K_S)$ for $\kappa=1$ is further reduced to
$8\times 10^{-5}$ and even $P(K_L\rightarrow K_L)$ becomes $0.05$,
while in the previous case it was $0.93$.

\begin{figure}
 \begin{center}
 \epsfig{file=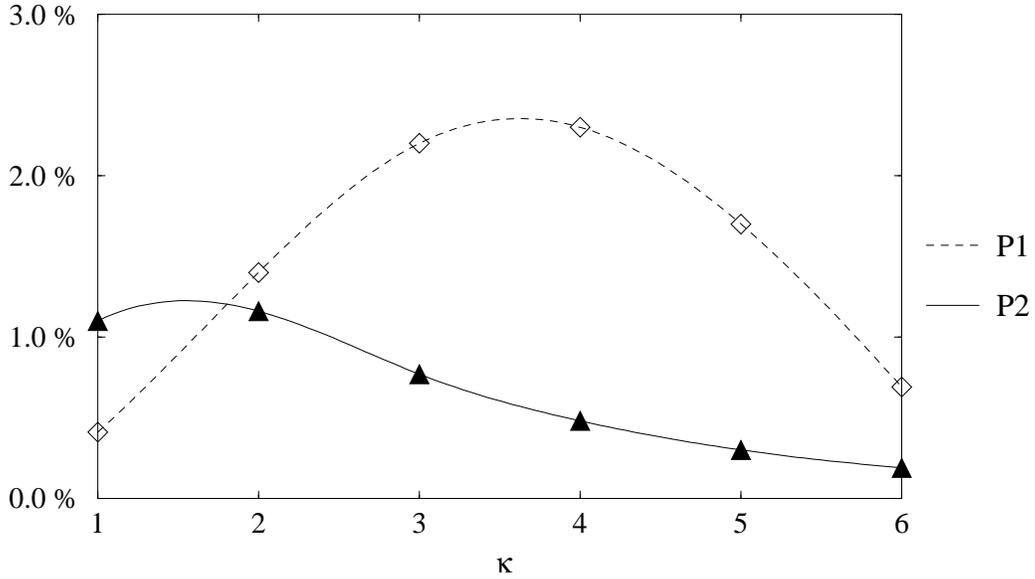,width=8cm,angle=270}
    \caption{$P_2$ is the $K_L\rightarrow K_S$ transition probability
    for $^{27}{\rm Al}\,-^{184}$W regenerator with
            $t_a=t_b=8\times10^{-11}$~s. $P_1$ is the same probability when
            neglecting all imaginary parts.}
    \label{fig:fig2}
  \end{center}
\end{figure}

Clearly, any measurable effect with kaons propagating in matter
requires times of the order of $\tau_S$. Unfortunately, for such
times, even the toy model without imaginary parts gives a
maximum value for $P(K_L\rightarrow K_S)$ of the order of 0.02 only.
Then, the damping due to the imaginary parts washes out the
effect. This is shown in Fig.~\ref{fig:fig2} where we compare for
the same $t_a$ and $t_b$ $P(K_L\rightarrow K_S)$ in the toy model
$(P_1)$ and for real kaons traversing a real $^{27}$Al -- $^{184}$W
regenerator.

\begin{figure}
 \begin{center}
 \epsfig{file=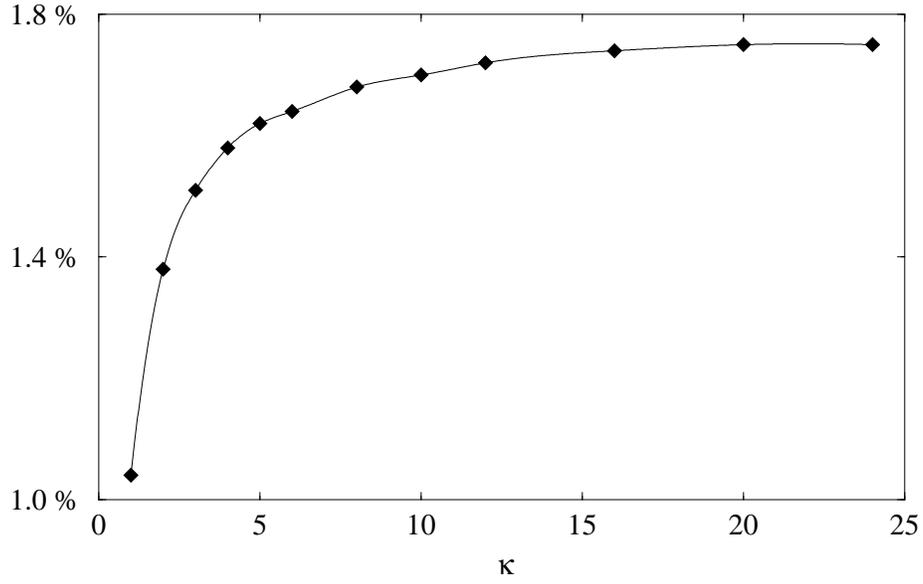,width=8cm,angle=270}
    \caption{$K_L\rightarrow K_S$ transition probability for
            $^{27}{\rm Al}\,-^{184}$W regenerator with
            $t_a=t_b=\frac{12}{\kappa}\times10^{-11}$~s.}
    \label{fig:fig3}
  \end{center}
\end{figure}

So far, in all cases that we have considered, the total time that
the particles spend in the regenerator, $t=\kappa(t_a+t_b)$,
increases linearly with $\kappa$. Obviously, after a few layers
most of the particles will disappear due to their decay or
absorption. Hence, it is interesting to examine another type of
experiment, where the total time $t$ is kept fixed, i.e. as
$\kappa$ increases the times $t_a$ and $t_b$ are proportionally
reduced. In Fig.~\ref{fig:fig3} we plot $P(K_L\rightarrow K_S)$ as
a function of $\kappa$ for this situation. For the beam velocity
that we are considering, $10^{-11}$~s corresponds to a pathlength
of the order of 1~mm in vacuum. Then from Fig.~\ref{fig:fig3}
one can see that a regenerator made of a 12~mm layer of $^{27}$Al
followed by another layer of 12~mm of $^{184}$W ($\kappa=1$) is
less efficient than another regenerator with four alternating
$^{27}$Al -- $^{184}$W layers of 6~mm each ($\kappa=2$).
Perhaps this effect is better illustrated if, instead of the
transition probability, we consider the decay of the kaons into a
final state $f$ after traversing the regenerator. From Eq.~(\ref{Kdecay})
one can calculate the time distribution $P(K_R(t)\rightarrow f)$ of the
final state $f$ after the kaon state initially produced as $K_L$ passes
through the regenerator and then spends outside it the proper time $t$
(which for simplicity we took equal to the proper time spent inside the
regenerator). The result is (e.g. ref.~\cite{BLS})
\begin{eqnarray}
P(K_R(t)\rightarrow f)&=&|{\langle f|T|K_S\rangle}|^2\,
|{\langle \tilde{K}_S|U_\kappa|K_L\rangle}|^2 \times \nonumber\\
&\times&\left[e^{ -\Gamma_S t} + \abs{v_f}^2 e^{-\Gamma_L
t}+2\abs{v_f}e^{-\Gamma t}\cos(\theta_f -\Delta m\cdot t)\right] \, ,
\end{eqnarray}
where
\begin{eqnarray}
v_f=\abs{v_f} e^{i\theta_f}\equiv \frac{\langle
\tilde{K}_L|U_\kappa|K_L\rangle} {\langle
\tilde{K}_S|U_\kappa|K_L\rangle}\eta_f \, .
\end{eqnarray}

\begin{figure}
 \begin{center}
 \epsfig{file=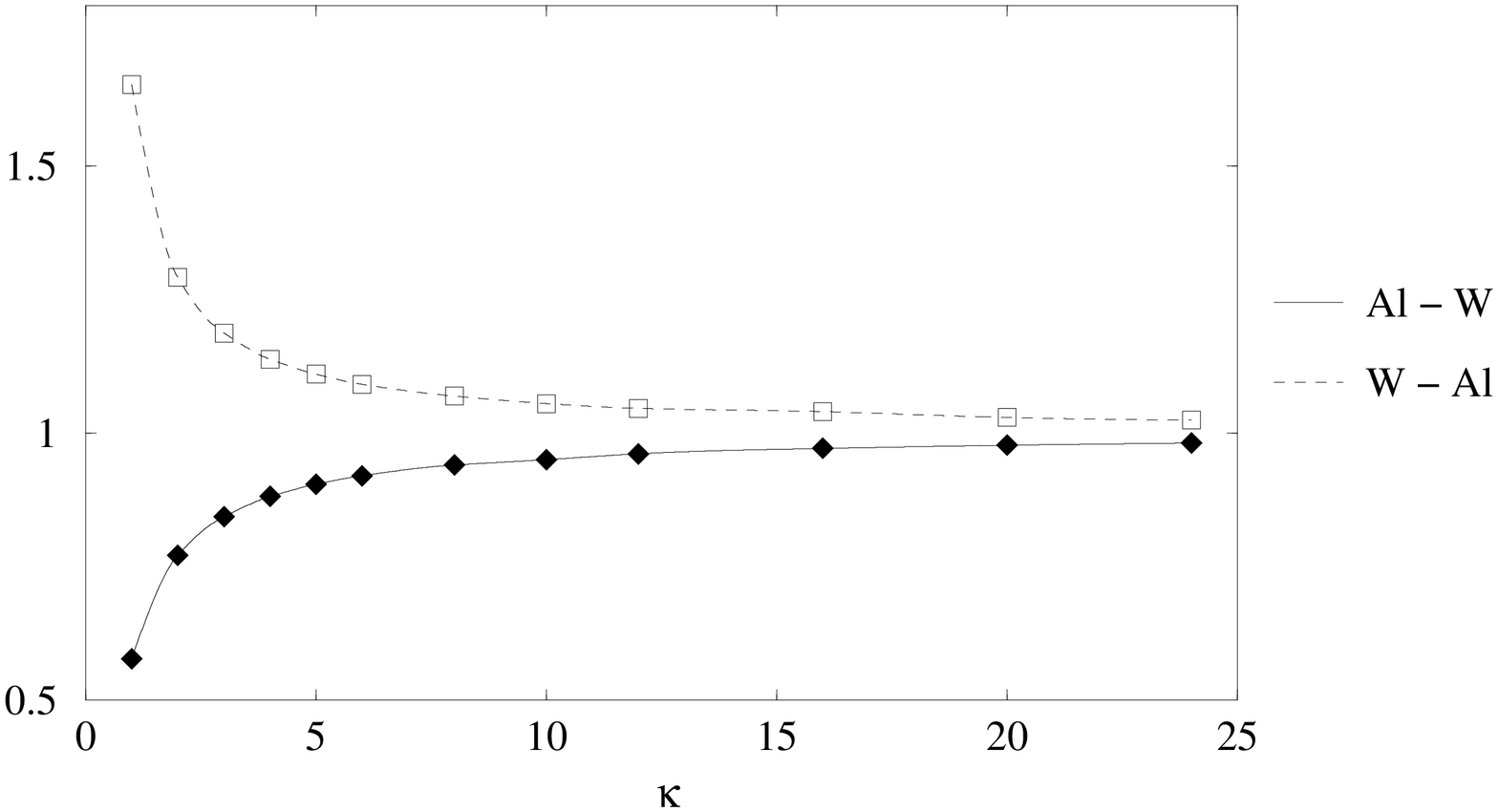,width=8cm,angle=270}
    \caption{$P(K_R\rightarrow\pi^+\pi^-)/(10^7{\rm s}^{-1})$ for
         two regenerators (Al-W or W-Al) with
            $t_a=t_b=\frac{12}{\kappa}\times10^{-11}$~s.}
    \label{fig:fig4}
  \end{center}
\end{figure}

In our example, shown in Fig.~\ref{fig:fig4}, we have assumed that one
measures the $\pi^+\pi^-$ final state. The magnitude and the phase of
$\eta_{+-}$ were taken from ref.~\cite{RPP}. The probability distribution
after passing $\kappa$ ($^{27}$Al-$^{184}$W) layer junctions increases
with $\kappa$. In the same Fig.~\ref{fig:fig4} we also plot the probability
distribution for a regenerator where the layers are in reverse order. In
this case $P(K_R\to f)$ decreases with $\kappa$, and both curves tend to a
common limit. This is easy to understand. As the number of layers increase
we are effectively approaching a ``mixed material'' with a density
that has the average density of aluminum and tungsten.
Since the regeneration effect is proportional to the density of the
regenerator, one can understand that $P(K_R\rightarrow\pi^+\pi^-)$
increases with $\kappa$ for the $^{27}$Al-$^{184}$W regenerator and
decreases in the $^{184}$W-$^{27}$Al case
\footnote{For $\kappa=1$ the regeneration in the $^{27}$Al-$^{184}$W case
is less efficient than in the $^{184}$W-$^{27}$Al one because a fraction
of $K_L$ decays in aluminum before they reach a more efficient regenerator 
-- tungsten.}. The variation with the order of the layers (notice that
their total number is fixed) is a nice example of quantum mechanics 
interference. In this problem, the evolution matrix for each individual
layer (cf. Eqs.~(\ref{Ua})-(\ref{Ub})) is an element of the $U(2)$ group.
Hence, the evolution for the total number of layers is, of
course, an element of $U(2)$. Since $U(2)$ is a non-Abelian group,
shuffling the layers one obtains a different evolution operator.
From this point of view, Fig.~\ref{fig:fig4} is a consequence of
the non-commutativity of the $U(2)$ group.

One should realize that the results shown in Fig.~\ref{fig:fig4}
are independent of the CP-violating parameter $\delta$. However,
it is possible to use this type of regenerators to measure
CP violation at the $\phi$-factories. To see how the effect
arises let us recall that the $\phi$-meson decays into the
antisymmetric combination
\begin{eqnarray}
\frac{1}{\sqrt{2}}\left[
|K^0({\bf p})\rangle|\bar{K}^0(-{\bf p})\rangle -
|\bar{K}^0({\bf p})\rangle |K^0(-{\bf p})\rangle \right]\, ,
\end{eqnarray}
where ${\bf p}$ denotes the momentum of the particle. We assume
that in the direction of $-{\bf p}$ we have a detector, called
"left", and in the direction of ${\bf p}$ another detector,
called "right". Both detectors measure muons from the
semileptonic decays of the kaons. These decay amplitudes are
\begin{eqnarray}
\langle \pi^-\mu^+\nu_\mu|T|K^0\rangle &= &A \, ,\\
\langle \pi^+\mu^-\bar{\nu}_\mu|T|\bar{K}^0\rangle
&= &A^\ast \, .
\end{eqnarray}
The kaons propagating to the right from the decay point have to traverse a
regenerator made of two layers of different materials $a$ and $b$. On the
other hand, the kaons that propagate to the left must traverse a similar
regenerator with two layers of the same width but in reverse order, $b$
followed by $a$. With this setting one can show that the amplitude to
detect in coincidence
two $\mu^+$ on both detectors is
\begin{eqnarray}
A\left(\pi^-,\pi^- \right)&=&\frac{A^2}{\sqrt{2}}\left(\, \langle
K^0|U_{ab}|K^0\rangle \langle K^0|U_{ba}-U_{ab}|\bar{K}^0 \rangle
\right.
+\nonumber\\ &+&x \left.\left[ \,\langle K^0|U_{ba}|\bar{K}^0 \rangle
\langle \bar{K}^0 |U_{ab}|K^0\rangle - \langle \bar{K}^0|U_{ba}|K^0\rangle
\langle K^0|U_{ab}|\bar{K}^0 \rangle\,\right]\,\right)\, \label{Amm}.
\end{eqnarray}
We have introduced the ratio
\begin{eqnarray}
x=\frac{\langle \pi^-\mu^+\nu_\mu|T|
\bar{K}^0\rangle} {\langle \pi^-\mu^+\nu_\mu|T|K^0\rangle}
\end{eqnarray}
between the $\Delta S=\Delta Q$ violating amplitude and the dominant
one. Experimentally, $x=\left[-2\pm 6+i(1.2\pm1.9)\right]\times
10^{-3}$ ~\cite{RPP}, which is consistent with zero; theoretically,
within the standard model one expects $x\sim 10^{-7}$.
Therefore in Eq.~(\ref{Amm})
we have neglected the term of order $x^2$. Finally, let us point
out that Eq.~(\ref{Amm}) goes trivially to zero when ¤a=b¤. This
is a simple consequence of the antisymmetry of the initial state.
With a similar notation one can obtain the amplitude for two
$\mu^-$ in coincidence. The result is
\begin{eqnarray}
A\left( \pi^+,\pi^+ \right) &= &\frac{(A^\ast)^2}{\sqrt{2}}
\left(\, \langle \bar{K}^0 |U_{ab}|\bar{K}^0 \rangle \langle
\bar{K}^0 |U_{ab}-U_{ba}|K^0\rangle \right. +\nonumber\\ &+&x^\ast
\left.\left[ \,\langle K^0|U_{ba}|\bar{K}^0 \rangle \langle
\bar{K}^0 |U_{ab}|K^0\rangle - \langle \bar{K}^0 |U_{ba}|K^0\rangle
\langle K^0|U_{ab}|\bar{K}^0 \rangle\,\right]\,\right)\, \label{App}.
\end{eqnarray}
The amplitude for a $\mu^+$ in the left detector and a $\mu^-$ in
the right detector is
\begin{eqnarray}
A\left( \pi^-,\pi^+ \right) &= &\frac{AA^\ast}{\sqrt{2}} \left(\,
\langle K^0|U_{ba}|\bar{K}^0 \rangle \langle \bar{K}^0 |U_{ab}|K^0\rangle
-\langle K^0|U_{ba}|K^0\rangle  \langle \bar{K}^0
|U_{ab}|\bar{K}^0\rangle
\right. +\nonumber\\ &+&x^\ast \left. \,\langle K^0|U_{ab}|K^0\rangle
\langle K^0|U_{ba}-U_{ab}|\bar{K}^0 \rangle + x\langle
\bar{K}^0 |U_{ab}|\bar{K}^0 \rangle \langle
\bar{K}^0 |U_{ab}-U_{ba}|K^0\rangle\,\right)\, \label{Amp}.
\end{eqnarray}
The other asymmetric amplitude $A(\pi^+,\pi^-)$ is
\begin{eqnarray}
A\left( \pi^+,\pi^- \right) &= &\frac{AA^\ast}{\sqrt{2}} \left(\,
\langle \bar{K}^0|U_{ba}|\bar{K}^0 \rangle \langle K^0
|U_{ab}|K^0\rangle
-\langle \bar{K}^0|U_{ba}|K^0\rangle  \langle K^0
|U_{ab}|\bar{K}^0\rangle
\right. +\nonumber\\ &+&x^\ast \left. \,\langle K^0|U_{ab}|K^0\rangle
\langle K^0|U_{ba}-U_{ab}|\bar{K}^0 \rangle + x\langle
\bar{K}^0 |U_{ab}|\bar{K}^0 \rangle \langle
\bar{K}^0 |U_{ab}-U_{ba}|K^0\rangle\,\right)\, \label{Apm}.
\end{eqnarray}
We shall now consider two asymmetries which can be measured in the 
two-arm experiments,
\begin{eqnarray}
R_1=\frac{ \abs{A(\pi^-,\pi^-) }^2 - \abs{A(\pi^+,\pi^+)}^2 }{
\abs{A(\pi^-,\pi^+)}^2}
\label{R1}
\\
R_2=\frac{ \abs{A(\pi^-,\pi^+) }^2 - \abs{A(\pi^+,\pi^-)}^2 }{
\abs{A(\pi^-,\pi^+) }^2 + \abs{A(\pi^+,\pi^-)}^2}\,, \label{R2}
\end{eqnarray}
and also an asymmetry which can be measured in the single arm experiments
of the CPLEAR type (see e.g. ref.~\cite{Benelli}),
\begin{eqnarray}
R_3=\frac{ \abs{A_{ab}(\pi^-) }^2 - \abs{A_{ba}(\pi^+)}^2 }{
\abs{A_{ab}(\pi^-) }^2 + \abs{A_{ba}(\pi^+)}^2} \label{R3}\,.
\end{eqnarray}
Here
\begin{equation}
A_{ab}(\pi^-)=A\left[ \langle K^0|U_{ab}|\bar{K}^0 \rangle +x\langle
\bar{K}^0|U_{ab}|\bar{K}^0\rangle\right]\,,
\end{equation}
\begin{equation}
A_{ba}(\pi^+)=A^\ast\left[ \langle \bar{K}^0|U_{ba}|K^0\rangle
+x^\ast\langle K^0|U_{ba}|K^0\rangle\right]\, .
\end{equation}
The ratios $R_1$ and $R_3$ are CP-asymmetric observables. They
depend on the intrinsic (i.e fundamental) CP violation
parameter $\delta$. Furthermore, since the regenerators are made
out of matter and not of equal amounts of matter and antimatter,
they are themselves CP-asymmetric and so induce a macroscopic,
extrinsic CP violation which in general contributes to both CP-violating 
observables, $R_1$ and $R_3$. However, interchanging the order of
the layers leads to a partial cancellation of the extrinsic CP
violating effects. For this reason the ratio $R_3$ is primarily
sensitive to the fundamental CP violation. In the limit $x=0$ the
cancellation of the extrinsic CP violation in $R_3$ is exact and this 
leads to the result $R_3\simeq 2\delta$. This is not so for $R_1$ which
does not vanish when $\delta=0$. The ratio $R_1$, for example, is 
normally of the order of unity as the extrinsic CP violation is of this 
order. For an aluminum-tungsten regenerator, using $x=0$,
$t_a=24\times10^{-11}$~s and $t_b=12\times 10^{-11}$~s, we find
$R_1 = 1.349$ for $\delta$ given in ref.~\cite{RPP}, whereas for
$\delta=0$ the corresponding value is $R_1=1.334$. One should
notice that $R_1$ is very sensitive to the imaginary parts of the
effective Hamiltonian. Switching off the imaginary parts of the
matter-induced potentials $V_i$ and $\bar{V}_i$ reduces $R_1$ by
about a factor of 200, while switching off the decay rates
$\Gamma_S$ and $\Gamma_L$ $R_1$ would reduce it by about a factor 
of 8. 
If all the imaginary parts are set equal to zero, $R_1$ is suppressed 
by a factor $2\times 10^{-4}$. 

The parameter $R_2$ may appear as a CP-violating observable too, but in
fact it is not. To see that one has to notice that CP transformation not
only interchanges particles with their antiparticles but also flips the
sign of all the momenta; for the two-arm setup under discussion this 
implies an additional interchange of the arguments of $A(\pi_i, \pi_j)$ 
so that $R_2$ is unchanged under the CP transformation $\delta\to
-\delta$. It has a moderate sensitivity to the imaginary parts of the 
effective Hamiltonian. For the same regenerator and $x=0$, we find
$R_2=-0.69$ with normal values of all the imaginary parts. Switching off 
$V_i$ and $\bar{V}_i$ reduces $|R_2|$ by about a factor of 2, while
switching off $\Gamma_S$ and $\Gamma_L$ would reduce it by about a factor
of 1.4. If all the imaginary parts are set equal to zero, $R_2$ goes to 
zero. 

Thus, by measuring $R_1$ and $R_2$ one can obtain an information on the
imaginary parts of the effective Hamiltonian of the $K^0\bar{K}^0$ system
in matter, and in particular on the imaginary parts of the
$K^0(\bar{K}^0)$-nucleon scattering amplitudes.


\vspace{0.25cm}
In conclusion, we have studied the effects of periodic matter in
kaon regeneration. Motivated by the possibility of the parametric
resonance in neutrino oscillations in matter we considered similar
effects in $K_L\rightarrow K_S$ transitions. Unfortunately, the large
$\Delta\Gamma$ and imaginary parts of the forward kaon-nucleon scattering
amplitudes prevent a sizable enhancement of the $K_L\rightarrow K_S$
transition probability (cf. Fig.~\ref{fig:fig2}). However, some interesting
effects can be produced using regenerators made of alternating layers of
two different materials. Despite the fact that the regenerator has a
fixed length one can obtain different values for the probability distribution
of the $K_L$ decay into a final state (cf. Fig.~\ref{fig:fig3} and
Fig.~\ref{fig:fig4}). Finally, we have pointed out that using a two-arm 
regenerator set up it is possible to measure the
imaginary parts of the $K^0(\bar{K}^0)$-nucleon scattering amplitudes
in the correlated decays of the $\phi$-resonance.
Combining the data of the single-arm regenerator experiments with direct
and reverse orders of the matter layers in the regenerator one can
independently measure the CP violating parameter $\delta$.

\vspace{0.25cm} \noindent{\it Acknowledgements.} We would like to
thank J.~P.~Silva for useful discussions and for a careful
reading of the manuscript. The work of E.A. was supported by
the ``Sonderforschungsbereich 375 f{\"u}r Astro-Teilchenphysik der
Deutschen Forschungsgemeinschaft''. P.K. wishes to thank the Jenny and
Antti Wihuri foundation and the Finnish Academy of Science and
Letters for financial support. This work was supported by
Funda\c{c}{\~a}o para a Ci{\^e}ncia e a Tecnologia through the
grants PRAXIS XXI/BPD/20182/99 and CERN/P/FIS/15183/99.


\pagebreak

\begin{thebibliography}{X}
\bibitem{renewed} Q. Y. Liu and A. Yu. Smirnov, Nucl. Phys. B 524
(1998) 505; Q. Y. Liu, S. P. Mikheyev and A. Yu. Smirnov, Phys. Lett.
B 440 (1998) 319; S. T. Petcov, Phys. Lett. B 434 (1998) 321;
E. Kh. Akhmedov, Nucl. Phys. B 538 (1999) 25; E. Kh. Akhmedov, A. Dighe,
P. Lipari and A. Yu. Smirnov, Nucl. Phys. B 542 (1999) 3.

\bibitem{ETC} V. K. Ermilova, V. A. Tsarev and V. A. Chechin, Kr. Soob.
Fiz. [Short Notices of the Lebedev Institute]  5 (1986) 26.

\bibitem{Akh1} E. Kh. Akhmedov, Yad. Fiz. 47 (1988) 475
[Sov. J. Nucl. Phys. 47 (1988) 301].



\bibitem{BLS} G. Castelo Branco, L. Lavoura and J.
P. Silva, {\it CP violation} (Oxford University Press, Oxford,
1999).

\bibitem{RPP} D. E. Groom {\em et al.}, Eur. Phys. J. C 15 (2000) 1.

\bibitem{reciprocalSilva} J. P. Silva, Phys. Rev. D 62 (2000) 11600P.

\bibitem{rescond} E. Kh. Akhmedov, in ref. \cite{renewed}.


\bibitem{scattamplitudes} R. Baldini and A. Nichetti, {\it $K_L$
Interactions and $K_S$ Regenaration in KLOE}. LNF-96-008-IR, 1996.

\bibitem{Benelli} A. Benelli, Nucl. Phys. Proc. Suppl.
75B (1999) 267.

\end{thebibliography}
\end{document}